\documentclass[3p,twocolumn]{elsarticle}

\journal{Physics Letters B}

\begin{document}

\begin{frontmatter}

\title{$P_{\gamma h}^{\alpha}$: a new variable for $\gamma / {\rm h}$ discrimination in large gamma-ray ground arrays}

\author[LIP,IST]{R. Concei\c{c}\~ao}
\author[LIP,IST]{B. S. Gonz\'alez}
\author[LIP,IST]{M. Pimenta}
\author[LIP,IST]{B. Tom\'e}


\address[LIP]{LIP, Av. Prof. Gama Pinto, 2, P-1649-003 Lisbon, Portugal}
\address[IST]{Instituto Superior T\'{e}cnico (IST), Universidade de Lisboa, Av. Rovisco Pais 1, 1049-001, Lisbon, Portugal}

\begin{abstract}

In this letter, a new strategy to enhance the discrimination of high energy gamma rays from the huge charged cosmic rays background in large cosmic rays ground arrays is presented. This strategy is based on the introduction of a new simple variable,  $P_{\gamma h}^{\alpha}$, which combines the probability of tagging muons and/or very energetic particles in each single array station. The discrimination power of this new variable, particularly important for and above multi-TeV energies, is illustrated for a few specific examples in the case of a hypothetical water Cherenkov detector cosmic ray array, both in the case of low and high particle stations occupancy. The results are very encouraging and hopefully will be demonstrated in the present and future gamma-ray Observatories.

\end{abstract}

\end{frontmatter}

%\linenumbers

\section{Introduction}

The capability to discriminate high energy gamma rays from the huge background of charged cosmic rays, hereafter nominated as $\gamma / {\rm h}$ discrimination, 
is presently one of the main challenges of ground-based gamma-ray observatories.
As an example, the signal/background ratio considering a Crab-like source,  integrated over 1 second within one square degree, in an effective area of $10\,000\,\rm{m^2}$, is, in the GeV-PeV energy region, of about $10^{-2}$- $10^{-3}$ \cite{Pimenta2018Astroparticle,LATTES}.
To handle these high background rates, several strategies have been used or proposed, based on the analysis of the distribution of the particles at the ground (steepness, compactness or bumpiness of the  Lateral Distribution Functions (e.g. \cite{LATTES,Greisen,Abeysekara_2017})
or, more globally, differences in the  detected shower patterns (e.g. \cite{Argo_pattern,assunccao2019automatic}). Around a few TeV, proton showers start to have a large number of muons that effectively can reach the ground. For the same reconstructed energy, gamma induced showers are expected to have much fewer muons than for proton-induced showers. Hence, the muonic content of the shower can be explored to achieve an excellent $\gamma / {\rm h}$ discrimination (see, for instance \cite{Blake_1995,Hayashida_1995,APEL2010202}).

The most common technique to identify muons in cosmic rays detectors is by placing sensitive tracking detectors under a reasonable number of equivalent radiation lengths to absorb the shower's electromagnetic (e.m.) component. This may be achieved  placing  the detectors  under earth ( e.g. \cite{1994NIMPA.346..329B,Aab_2021,LHAASO_muon}), water or ice (e.g. \cite{Milagro}), concrete or some other inert material.  Another option is to design detectors with at least two active layers, where the first one(s) act for this purpose as an absorber (e.g. \cite{MARTA}, \cite{DLWCD_Antoine}). 
An alternative/complementary approach is to have, at the station level, several signal sensors (e.g. PMTs) and explore time and/or intensity differences between them (e.g. \cite{zuniga2017detection,Gonzalez_2020,Borja4PMTs}).

Nevertheless, in the end, the capability to tag, in each individual station, the presence of one or more muons is a compromise between purity (low or negligible punch-through) and efficiency, which may be translated in a probability, ${P_{\mu,i}}$. At the event level, the question would be then how to handle such probabilities in order to reach the needed $\gamma / {\rm h}$ discrimination ensuring a reasonable efficiency. 

In this letter, we propose a new, very simple variable, $P_{\gamma h}^{\alpha}$, that combines efficiently the individual ${P_{\mu,i}}$ and avoids/minimises the need of additional fiducial cuts. This quantity is particularly important when performing the discrimination at multi-TeV energies and above. The document is organised as follows: the discriminator is presented in Section~\ref{sec:Pgh}; in Section~\ref{sec:simulation} the simulation setups are described, followed by the presentation of the obtained results (Sec.~\ref{sec:results}), including a discussion of these. The paper ends with a short summary.

%The definition of the new variable is presented in the next section, while its discrimination power is illustrated in the subsequent sections, both in the case of low and high particle stations occupancy\footnote{in this context occupancy refers to the number of secondary shower particles hitting one single station.}. The used working case is one of the layout proposals for the future Southern Wide-field Gamma-ray Observatory (SWGO) \cite{SWGO}.
%It is based on small water Cherenkov detectors (WCD) organised in a compact array surrounded by a sparse array with variable fill factor as a function of the distance to the centre of the array.

%Afterwards, the $P_{\gamma h}^{\alpha}$ variable may be included in a pattern recognition analysis that would ensure the ultimate background rejection power for a target signal efficiency. This second level of analysis is out of the scope of the present letter.

%\section{The $P_{\gamma h}^{\alpha}$ variable}
\section{Probability of muon detection for $\gamma/h$ discrimination}
\label{sec:Pgh}

It has been shown in~\cite{Borja4PMTs} that the sum of the probability of having a muon in a water Cherenkov station, $P_{\gamma h}$, obtained through the analysis of the photomultipliers (PMTs) signal time trace, can be used to discriminate between gamma and hadron induced showers. This discrimination was demonstrated for showers with energies of about $1\,$TeV. As the shower energy increases, the number of muons that might hit a WCD station will increase. However, the number of stations touched by the electromagnetic shower component and without muons will grow even faster, leading to an artificially larger $P_{\gamma h}$.
In this article, we argue that this effect can be mitigated without recurring to any type of cuts while maintaining an excellent gamma/hadron discrimination capability, with the introduction of a new variable, $P_{\gamma h}^{\alpha}$.

\begin{figure}[!t]
 \centering
\includegraphics[width=0.8\linewidth]{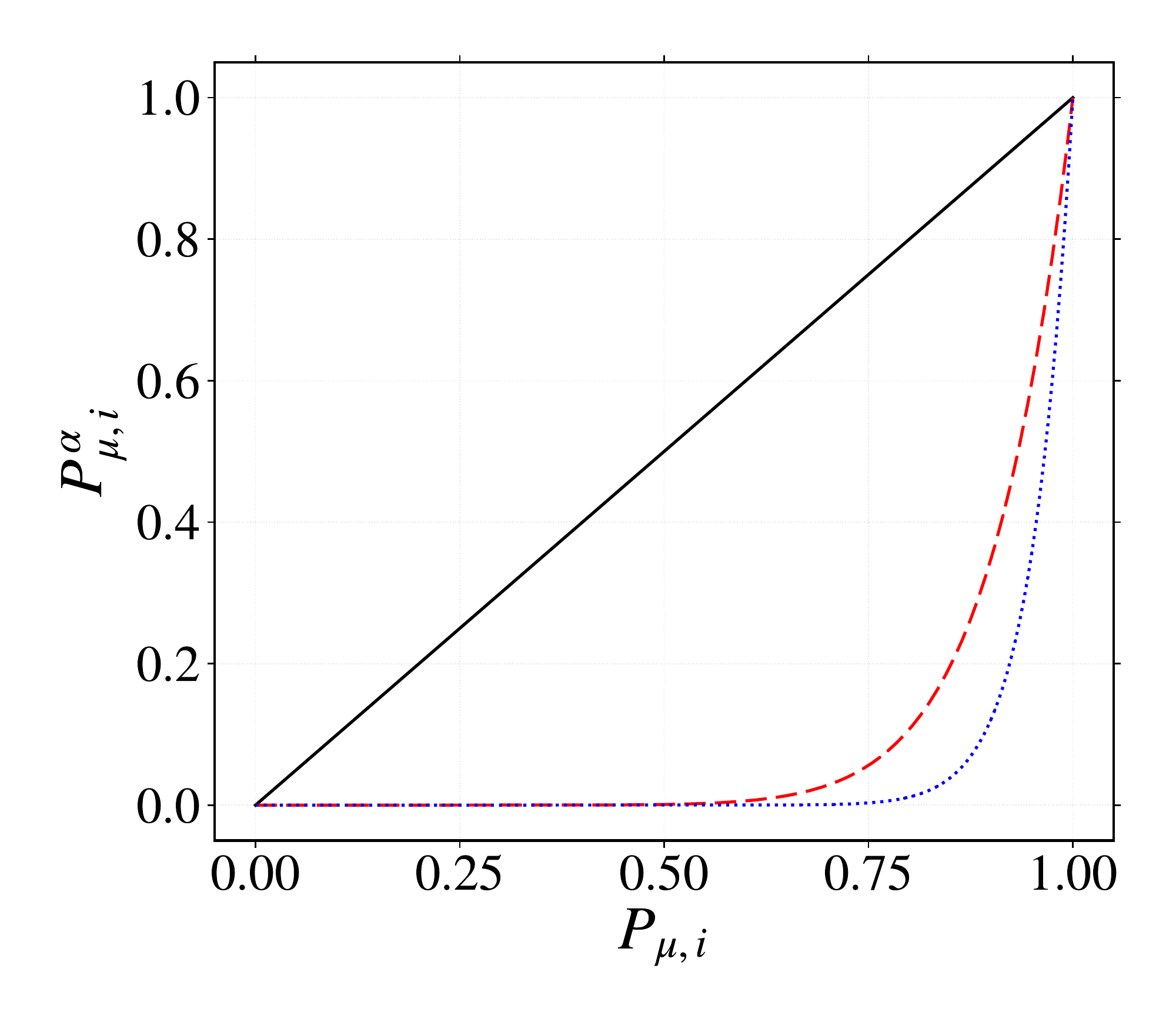}
\caption{\label{fig:alpha} 
${P_{\mu,i}}^{\alpha}$ as a function ${P_{\mu,i}}$ for different values of ${\alpha}$. 
The full black, dashed red and point blue lines  corresponds respectively to  ${\alpha}$ values of  1, 10, 20. 
See text for a discussion on the role of the constant ${\alpha}$ in the design of the global event $P_{\gamma h}^{\alpha}$ discriminant variable.
 }
\end{figure}

The proposed new variable is defined as:

\begin{equation}
P_{\gamma h}^{\alpha}= \sum_{i}^{n} {P_{\mu, i}}^{\alpha} 
\label{eq:Pgh}
\end{equation}

where $P_{\mu,i}$ is the probability of a station being hit by a muon, $n$ the number of active stations in the shower event, and $\alpha$ a parameter that maximises the separation between gamma and proton-induced showers.
For $\alpha$ = 1 $, P_{\gamma h}^{\alpha}$ is just the sum of the probabilities of all individual stations. On the other hand, the setting of $\alpha > 1$  enhances the relative weight of the stations where ${P_{\mu,i}}$ is close to 1. Indeed, as it is shown in Fig.~\ref{fig:alpha}, the use of high $\alpha$ values decreases dramatically the contributions to the $P_{\gamma h}^{\alpha}$ discrimination variable of the stations with lower probabilities of having been hit by a muon. The precise value of $\alpha$ to be used in each specific analysis should be optimised according to the desirable requirements taking into account the signal and background efficiency curves as well as the total number of stations foreseen as being hit by muons.

\section{Simulation Setup}
\label{sec:simulation}
To verify the discrimination capability of the proposed variable, $P_{\gamma h}^\alpha$, two distinct simulation frameworks are used. One is aiming at low particle station occupancy\footnote{In this context, occupancy refers to the number of particles hitting one single station. These two cases are distinguished as the strategy to identify muons should be distinct.} (shower energies of $\sim 50\,$TeV) while the other is used for high occupancy ($\sim 500\,$TeV). The simulation sets can be summarised in the following way:
\begin{itemize}
    \item A compact array of stations covering an area of $80\,000\,{\rm m^2}$ with an end-to-end simulation. The probability of having a muon in a station is extracted through the analysis of the PMT signal time trace using a Machine Learning algorithm.
    \item The same compact array but now surrounded by a sparse array with an area of $1\,{\rm km^2}$ and $1\%$ fill factor, simulated recurring to a fast simulation. The shower muon content is obtained via statistically discounting the electromagnetic component recorded at the ground.
\end{itemize}

For the first set of simulations, we consider a ground array of small-WCD stations where the light signal is collected by four PMTs placed at the bottom of the station, as described in references \cite{Borja4PMTs,ICRC_4PMT}. The considered array covers an area of $80\,000\,{\rm m^2}$ and has a fill factor of $\sim 85\%$ ($5\,720$ stations). 
The simulations of extensive air showers were generated using CORSIKA (version 7.5600) \cite{CORSIKA}. For this set, only protons with primary energy between $40$ - $63\,$TeV and a zenith angle between $5^\circ$ - $15^\circ$ were used. The discrimination capability is assessed by comparing proton-induced shower events with the same events but removing stations with muons. As we want to evaluate the ability to discriminate between showers using the number of muons in the event, this test provides a conservative assessment.

A total of $3\,686$ showers\footnote{The number of showers used to train and test the CNN were $2\,000$ and $1\,686$, respectively.} have been generated with a uniform logarithm of energy distribution, leading to a mean energy of $\sim 50\,$TeV. The detector response was simulated with the Geant4 toolkit \cite{agostinelli2003geant4,Geant4_2006,Geant4_2016}, and the shower core was distributed uniformly in a circular area centred in the centre of the array and with a radius of $50\,$m.

%The probability ${P_{\mu,i}}$ of tagging at least one muon with a signal greater than 300 p.e. was determined by exploring both the temporal and spatial asymmetries between the PMT signal time traces. These traces were analysed through a Convolutional Neural Network (CNN) \cite{BorjaOtherPaper}. 

The probability ${P_{\mu,i}}$ of tagging at least one muon with a signal greater than $300\,$p.e. (the minimum signal of vertical muons) was determined through a Convolutional Neural Network (CNN) similar to the one used in \cite{Borja4PMTs} at lower energies ($\sim 1$ TeV). To explore both the temporal and spatial features, the model receives as inputs: the normalised signal time trace of each PMT; the integral of each PMTs signal time trace; the amount of Cherenkov light measured in the WCD; and the normalised integral of each PMTs signal time trace. At this energy range ($\sim 50$ TeV), a different structure of the input signals was found to be better to deal with the overwhelming electromagnetic contamination. A tensor with dimensions $[3,3,30]$ is built using the signals (the empty spaces are filled with zeros). The $3 \times 3$ matrix allows to mimic the PMT spatial position while the remaining dimension contains the time evolution of the PMT signal time trace, being each entry a time bin with a width of $1\,$ns. Then, a 3-dimensional CNN (3D-CNN) is used to convolve over this tensor.  

\hfill \break

For the second set of simulations, the distributions were obtained using a fast and flexible simulation framework~\cite{Fast_sim}. This framework receives as input either CORSIKA showers or toy events generated as the superposition of single photons, electrons and muons, taking into account the energy spectrum of these single particles and the total mean electromagnetic and muonic particle lateral distributions obtained from dedicated CORSIKA simulations. The response of the detector was included as parameterised functions obtained by injecting single particles in the WCD and simulating it with Geant4. The configuration of the ground array that has been tested is composed of a compact array with an area of  $80\,000\,{\rm m^2}$ and a fill factor of $\sim 80\%$ surrounded by a sparse array with an area of $1\,{\rm km^2}$ and a fill factor of $\sim 1\%$. The WCD stations were similar to those described in the previous section, but only one PMT per station was placed at the bottom of the tank. 

\section{Results}
\label{sec:results}
\subsection{$P_{\gamma h}^\alpha$ at low particle stations occupancy}

\begin{figure}[!t]
 \centering
\includegraphics[width=0.8\linewidth]{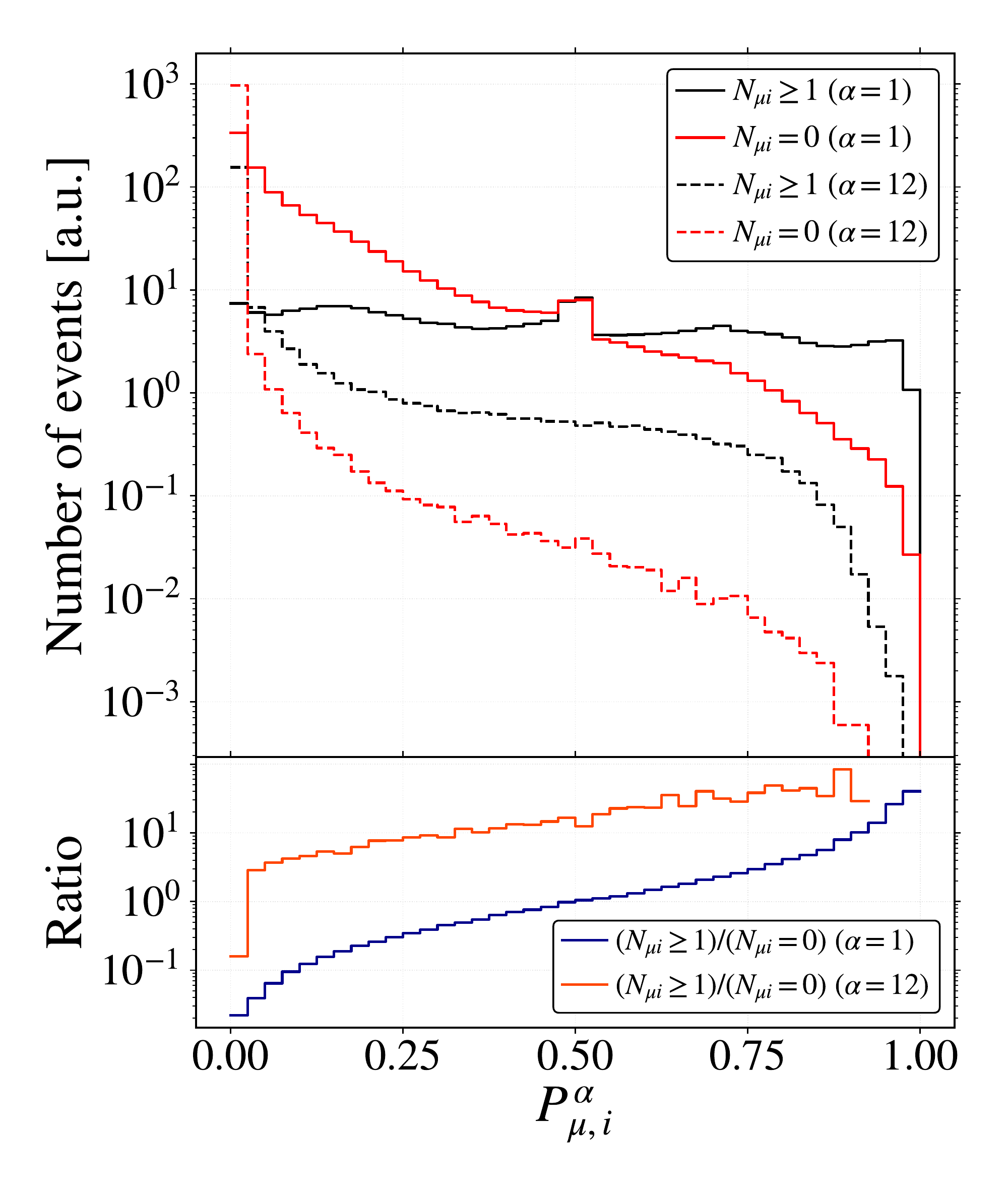}
\caption{\label{fig:alphaStation} 
(top) Mean number of stations per shower as a function of ${P_{\mu,i}}^{\alpha}$ for all the stations with distances from the shower core greater than $50\,$m. Stations with muons are represented by black curves and in red if there is no muon present. The full (dashed) lines are the ${P_{\mu,i}}^{\alpha}$ with $\alpha = 1$ ($\alpha = 12$). (bottom) Ratio between stations with and without muons for $\alpha = 1$ (blue) and $\alpha = 12$ (orange).
 }
\end{figure}

The impact of using $\alpha > 1$ at the station level can be seen in Fig.~\ref{fig:alphaStation} where the ${P_{\mu,i}}^{\alpha}$ distribution is shown for $\alpha=1$ and $\alpha=12$ for stations with and without the presence of muons. 
From these plots, it can be seen that whenever a cut in $P_{\gamma h}^{\alpha}$ is used to tag muons,  the choice of $\alpha > 1$ reduces the relative importance of the large number of stations without muons in a gamma or hadron high energy event. Such enables a lower cut which effectively increases the efficiency to tag stations with muons.

\begin{figure*}[!t]
 \centering
\includegraphics[width=0.4\linewidth]{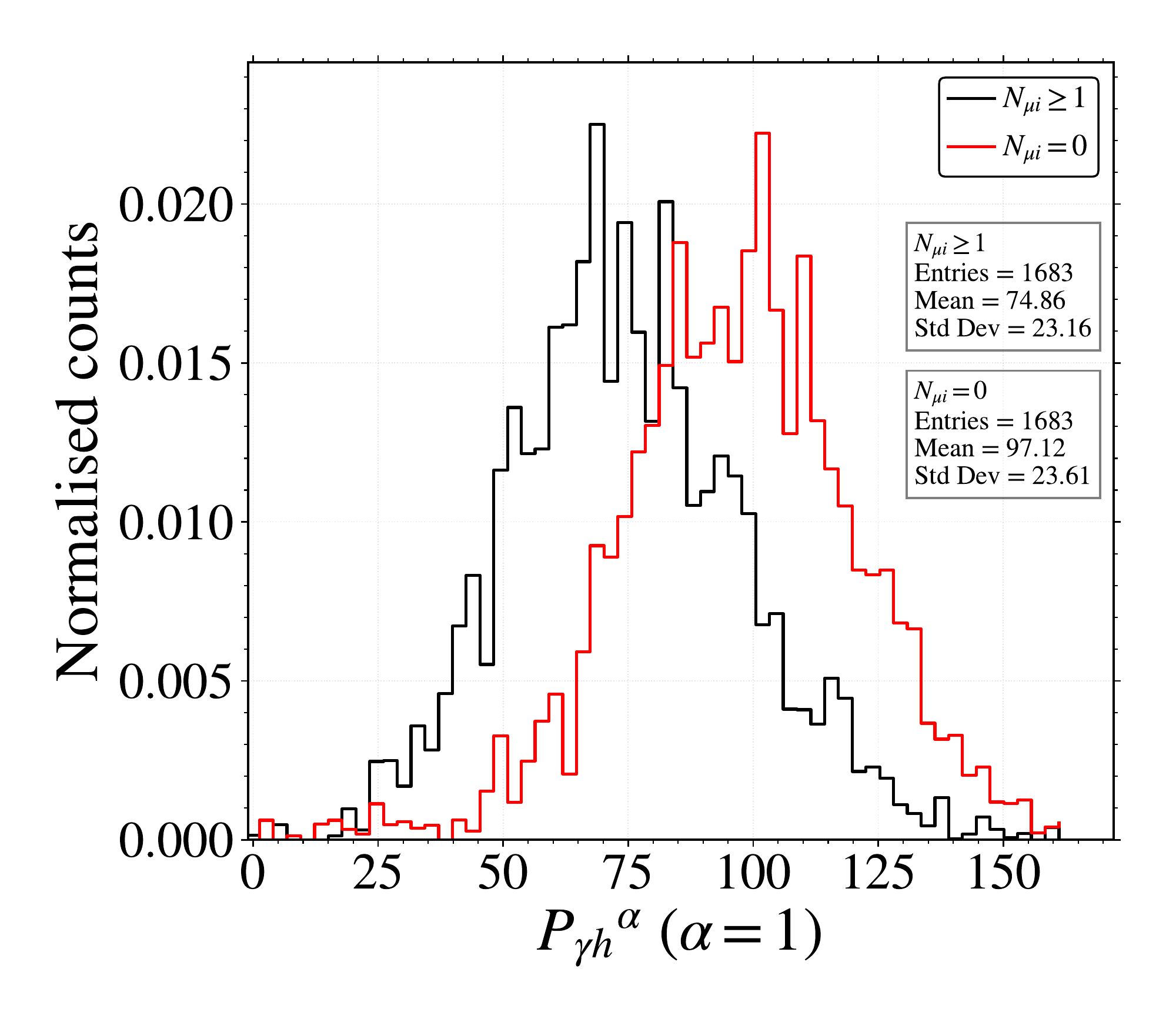}
\includegraphics[width=0.4\linewidth]{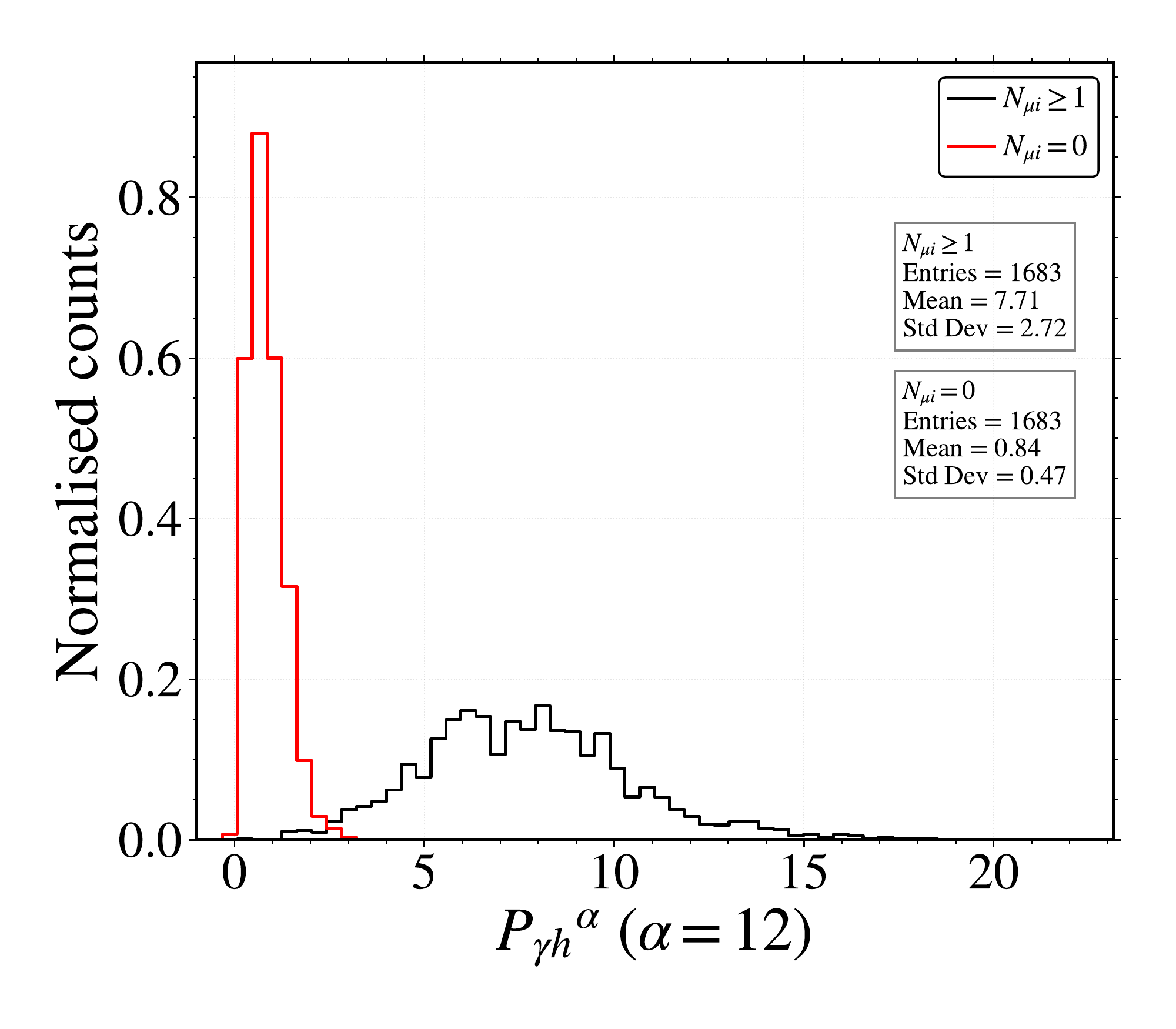}
\caption{\label{fig:PghCNN} 
 Distributions of the variable $P_{\gamma h}^{\alpha}$ at low particle stations occupancy events setting $\alpha $ to 1 (left) and 12 (right) in stations without muons (red lines) and at least one muon (black line). 
 }
\end{figure*}

The distributions of the variable $P_{\gamma h}^{\alpha}$ setting $\alpha $ to 1 and 12 were then obtained and are represented in Fig.~\ref{fig:PghCNN}, both for the stations without muons and for the stations hit at least by one muon. Only WCDs placed at a distance greater than $50\,$m from the shower core were considered to limit the electromagnetic contamination in the station, allowing for the identification of muons. Notice that after this cut, the number of stations with no muons with respect to stations with muons is still extremely high. The value of $\alpha $ = 12 was found to maximise the ratio between the signal and the square root of the background $(S/ \sqrt{B})$ for a signal efficiency $S=0.6$. 

The use of an optimised $ \alpha $ value has thus greatly increased the capability to identify the stations hit by muons which, in a full data analysis, would be translated in a significant increase of the $\gamma / {\rm h}$ discrimination power, at these energies and detector configuration.

\subsection{$P_{\gamma h}^{\alpha}$ at high stations occupancy}

At very high energies, the fluxes of high energy gamma and charged cosmic rays decrease dramatically, but, on the other hand, the density of particles at the ground for each particle shower increases even faster than the primary cosmic-ray energy. The detector ground array has thus to cover larger areas, but the array fill factor can be much smaller (sparse arrays).

In stations where the deposited electromagnetic energy is higher than some hundreds of MeV, the identification of muons using temporal and spatial asymmetries between the PMTs signal time traces is no longer efficient.
However, statistical methods, similar to the ones used in IceTop/IceCube~\cite{icetop}, may be used. The proposed muon tagging is based on the fact that in a station hit by one or more muons, or by one very energetic particle, the total signal detected is, on average, greater than the signal registered in one station located at a similar distance to the shower core in the case of one gamma event with a similar energy at the ground. 

This signal excess in station $i$ at distance $r$ and with total signal $S_i(r)$ may thus be quantified by :

\begin{equation}
    n_i = \frac{S_i(r) - S_{em}(r)}{\sigma_{em}(r)}
    \label{eq:1}
\end{equation}

\noindent where $S_{em}(r)$ and $\sigma_{em}(r)$ are respectively the expected electromagnetic signal in station $i$, and  the standard deviation of its expected fluctuations, for a gamma event. Whenever, $n_i < 0$, $n_i$ is set to $0$. $S_{em}(r)$ and $\sigma_{em}(r)$ may be evaluated from data, in events classified with a high probability to be gamma events and with a similar  energy at ground. However,
in the present work both $S_{em}(r)$ and $\sigma_{em}(r)$ were taken from CORSIKA simulations.

Then, for large $n$, the probability of a fluctuation of the electromagnetic signal, $P_{em, n} \equiv P(S_{i} \ge S_{em} + n \sigma_{em}$), can be approximated by

\begin{equation}
P_{em,n} = \frac{e^{-n^2/2}}{2 n \sqrt{\pi/2}}.
\end{equation}

where it is assumed that $P_{em,n}$ behaves as a normal distribution. The  probability to correctly identify a station as having at least one muon or one  high energy particle can then be defined as: 

\begin{equation}
 P_{\mu,i} = 1 - P_{em,n_i}. 
\end{equation}

\begin{figure}[!t]
 \centering
\includegraphics[width=0.8\linewidth]{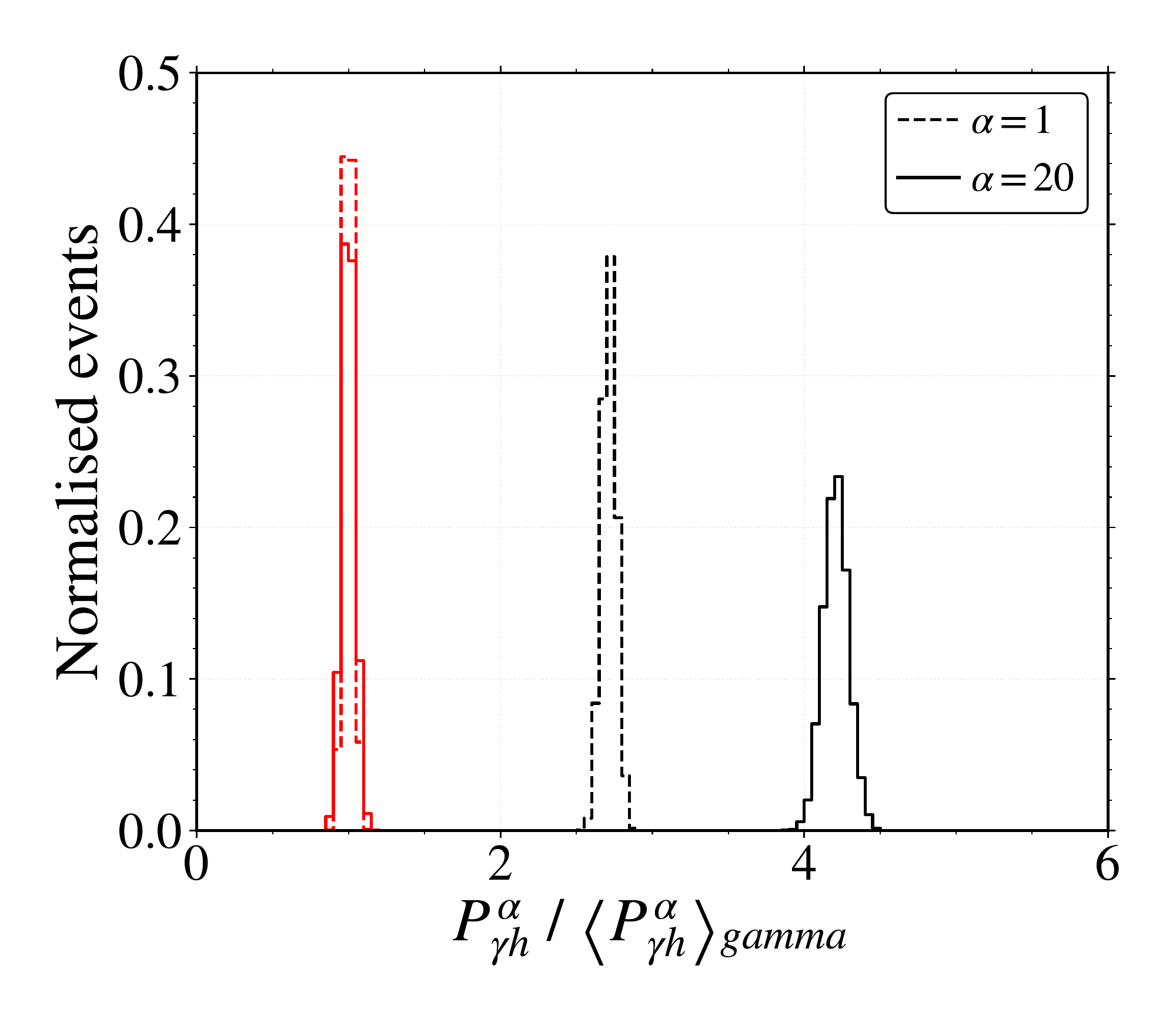}
\caption{\label{fig:Pgh_High} 
 Distributions of the variable $P_{\gamma h}^{\alpha}$ at high particle stations occupancy events setting $\alpha $ to 1 (dashed lines) and 20 (full lines) for photon (red) and proton (black) showers. The distributions are normalised to the mean values of the $P_{\gamma h}^{\alpha}$ distribution in the case of $\gamma$
 showers.
 }
\end{figure}

The variable $P_{\gamma h}^{\alpha}$ defined in equation  \ref{eq:Pgh} 
was shown  to be very effective in having a very high $\gamma / {\rm h}$ 
discrimination power at energies of few hundreds of TeV or higher, using $\alpha \simeq 20$.
Indeed, in Fig.~\ref{fig:Pgh_High} are shown the distributions of the variable $P_{\gamma h}^{\alpha}$ setting $\alpha $ to 1  and 20  for $\gamma$  and proton-induced showers with a primary energy at ground equivalent to a 500 TeV $\gamma$-shower and with the core generated uniformly in a ring centred in the centre of the array and inner and outer radius respectively of 240 m and 260 m. To make clear the impact on the expected separation between $\gamma$  and proton-induced showers, the distributions were normalised to the expected mean values of the relevant $\gamma$ distribution. 
In this case, the relative separation between proton and $\gamma$ distributions increases by a factor of 2, setting $\alpha$ = 20. 

Moreover, the resilience to the presence of tails in the $P_{\mu,i}$ distribution, expected in real life, is much higher whenever $\alpha > 1$. In fact, these tails may reflect simplified simulation frameworks where the signal fluctuations were not fully described (for instance,
 charged particle specific trajectories crossing the PMTs photo-cathodes, or smaller muon paths inside the stations; or fluctuations in the light sensor gains) 
which tends to populate the intermediate region of the $P_{\mu,i}$ distributions, whose contributions, as it was referred to in section \ref{sec:Pgh},  
decreases  for higher $\alpha $ values.

\section{Conclusions}

In this letter, we show that a very simple $\gamma / {\rm h}$ discrimination variable, $P_{\gamma h}^{\alpha}$, can be built using the individual ${P_{\mu,i}}$. The probability of having a muon in the station,  ${P_{\mu,i}}$, can be weighted by an exponent $\alpha$ maximising the discrimination power and minimising the need for additional cuts. The proposed method is applied to both low and high energy showers, with notable results demonstrating the power of this approach.

\section*{Acknowledgements}
We would like to thank to Alessandro De Angelis and Alan Watson for carefully reading the manuscript and useful comments.
The authors thank also for the financial support by OE - Portugal, FCT, I. P., under project PTDC/FIS-PAR/4300/2020. R.~C.\ is grateful for the financial support by OE - Portugal, FCT, I. P., under DL57/2016/cP1330/cT0002. B.S.G. is grateful for the financial support by grant LIP/BI - 13/2021 under project PTDC/FIS-PAR/4300/2020, and the FCT PhD grant PRT/BD/151553/2021 under the IDPASC program. 

\bibliography{references}

\end{document}